\documentclass[aps,preprint,groupedaddress]{revtex4}
\usepackage{epsfig,amsbsy,bm}
\newcommand{\eref}[1] {(\ref{#1})}
\newcommand{\Eref}[1] {Eq.~(\ref{#1})}
\newcommand{\Fref}[1] {Figure \ref{#1}}
\newcommand{\isum}%
{\mathop{\hbox{$\displaystyle\sum\kern-13.2pt\int\kern1.5pt$}}}

\begin{document}
\bibliographystyle{apsrev}
\baselineskip = 8mm

\title{
Two-photon double ionization of helium in
the region of photon energies 42-50~eV.}

\author{I. A. Ivanov\footnote[1]{Corresponding author:
Igor.Ivanov@.anu.edu.au}\footnote[2]{On leave from the Institute of
Spectroscopy, Russian Academy of Sciences}
 and A. S. Kheifets}
\affiliation
{Research School of Physical Sciences and Engineering,
The Australian National University,
Canberra ACT 0200, Australia}

\date{\today}
\begin{abstract}
We report the total integrated cross-section (TICS) of two-photon
double ionization of helium in the photon energy range from 42 to
50~eV. Our computational procedure relies on a numerical solution of
the time-dependent Schr\"odinger equation on a square-integrable basis
and subsequent projection of this solution on a set of final states
describing two electrons in continuum.  Close to the threshold, we
reproduce results previously known from the literature. The region
$47-50$ eV seems to have been previously unexplored.  Our results
suggest that TICS, as a function of the photon energy, grows
monotonously in the region $42-50$~eV.  We also present fully resolved
triple differential cross sections for selected photon energies.
\end{abstract}

\maketitle

\section{Introduction}
\label{S1}

Multi-photon atomic ionization resulting in ejection of a single
electron, as well as other single active electron phenomena in intense
laser fields, are relatively well understood by now \cite{PKK97}. In
contrast, strong field ionization with several active electrons
involved is a much more challenging problem in which the highly
nonlinear field interaction is entangled with the few-body correlated
dynamics \cite{BDM05}. The two-photon double-electron ionization
(TPDI) of helium is the archetypal reaction of this kind. Even for
this simplest many-photon many-electron process, non-perturbative
treatment of the external field is essential as well as a proper
account of correlation in the two-electron continuum. Neglect of
either aspects of TPDI results in a gross
failure.  In Ref.~\cite{CP02a}, for instance, it was demonstrated that
a perturbative treatment of the external field in this process can
lead to an order of magnitude error in the cross-sections even for
relatively mild fields.

Because of canonical importance of the TPDI of He, a number of
theoretical methods have been developed and applied to this problem
recently.  Among them are the so-called many-electron many-photon
theory \cite{mn1,mn2}, the $R$-matrix Floquet approach
\cite{FdH03}, and various time-dependent approaches
\cite{kamst,pbl,tdse1,tdse2,tdse3,PR98a,CP02a}.
These studies allowed to achieve considerable progress in theoretical
modelling of TDPI in helium. As far as total ionization cross section
(TICS) is concerned, the region of the photon energies from the
threshold (38.5~eV) to 47~eV is well understood. Various methods, such
as the time-dependent close-coupling (TDCC) approach
\cite{hcc2005,CP02a,cp2004} and the R-matrix Floquet method
\cite{FdH03}, gave results which lie sufficiently close to each other,
and which indicate that in this region of the photon energies TICS is
a monotonously growing function of the energy.  In Ref.~\cite{har} the
presence of a maximum of TICS in the vicinity of 42~eV was
reported. For larger energies, the authors found that TICS starts
decaying monotonously.  Overall shape of TICS, as a function of the
photon energy, was found to be very similar to that of single-photon
double ionization. However, this finding contradicts to other reports
which indicated no maximum anywhere below 47~eV.

In the present work, we report the behavior of TICS of TPDI of helium
at larger energies from 47 to 50~eV. This photon energy range seems to
be unexplored up to now. Our results indicate that TICS continues
to grow in this region of energies.

As a computational tool, we used a method which we proposed recently
for single photon double ionization studies \cite{tdsep}.  The method
is based on a numerical integration of the time-dependent
Schr\"odinger equation (TDSE) with subsequent projection of the
solution on a set of the field-free final states of the helium atom
with both electrons in continuum.  Accurate description of these
states is by itself a rather complicated problem.  In
Refs.~\cite{bach2,BFHMF05}, inter-electron correlations in the final
state was taken into account perturbatively.  One can also
address this problem using the exterior complex scaling method
\cite{ecs1,ecs2,ecs3} or using the complex Sturmian basis
\cite{csb}. The hyperspherical R-matrix method with semiclassical
outgoing waves \cite{MSK00} and various implementations of the
close-coupling method \cite{cpl2002,cp2004,B94,bstel} were also used.

In our earlier work \cite{tdsep}, we proposed to use the so-called
convergent close-coupling (CCC) expansion \cite{FB97tr} to describe
the field-free two-electron continuum in conjunction with solution of
TDSE.  In that paper we considered effect of the external DC electric
field on the single-photon double-electron ionization cross
section. In the present work, we apply this method for the study of
two-photon double electron ionization of helium.

The paper is organized as follows. In the next section we give an
outline of the theoretical procedure.  Then we discuss the results we
obtained for the integrated and fully differential cross sections of
TPDI of helium.

\section{Theory.}
\label{S2}

Detailed description of our method can be found in
Ref.~\cite{tdsep}. We shall present here only a brief description of
the computational procedure.
At the first step we solve numerically the TDSE for the helium atom in
the presence of the external ac field:
\begin{equation}
\label{TDSE}
i \ \partial \Psi/ \partial t=\hat H \Psi,
\end{equation}
where:
\begin{equation}
\hat H=\hat H_0+
\hat V_{12}+
\hat H_{\rm int}(t),
\label{ham}
\end{equation}
where the non-interacting Hamiltonian and the Coulomb interaction are,
respectively,
\begin{equation}
\hat H_0={{\bm p}_1^2\over 2}+
{{\bm p}_2^2\over 2}
-{2\over r_1}-{2\over r_2},
\label{h0}
\end{equation}
\begin{equation}
\hat V_{12}={1\over|{\bm r}_1-{\bm r}_2|} \ .
\label{v12}
\end{equation}
The interaction with the external ac field is written in the length
gauge:
\begin{equation}
\hat H_{\rm int}(t)=f(t)
({\bm r}_1+{\bm r}_2)\cdot
{\bm F}_{\rm ac}\cos{\omega t}
\label{hint}
\end{equation}
Here $f(t)$ is a smooth switching function which is chosen in such a
way that the amplitude of the field remains constant during the time
interval $(T,4T)$, where $T=2\pi/\omega$ is a period of the ac
field. This field is ramped on and off smoothly over one ac field
period. The total duration of the atom-field interaction is therefore
$T_1=6T$.

The solution of the TDSE is sought in the form of
expansion on a square-integrable basis
\begin{equation}
\Psi({\bm r}_1,{\bm r_2},t)=\sum\limits_{j}
a_j(t) f_{j}({\bm r}_1,{\bm r_2}) .
\label{exp}
\end{equation}
Here
\begin{equation}
f_{j}({\bm r}_1,{\bm r}_2)=
\phi^N_{n_1l_1}(r_1) \phi^N_{n_2l_2}(r_2) \
|l_1(1)l_2(2)\ L \rangle,
\label{basis}
\end{equation}
where notation $|l_1(1)l_2(2)\ L \rangle$ is used for
bipolar harmonics.
The radial orbitals in \Eref{basis} are the so-called pseudostates
obtained by diagonalizing the He$^+$ Hamiltonian in a Laguerre basis
\cite{B94}:
\begin{equation}
\langle\phi^N_{nl}|{\hat H}_{\rm{He}^+}|\phi^N_{n'l'}\rangle=E_i
\delta_{nn'}
\delta_{ll'}
\end{equation}

In the present work, we consider electric field of the order of $0.1$
a.u. corresponding to $3.5\times 10^{14}$~W/cm$^2$ intensity.  For
this, not very high intensity, we can retain in the expansion
~\eref{exp} only the terms with total angular momentum $J=0-2$. To
represent each total angular momentum block, we proceed as follows.
For all $S$, $P$, $D$ total angular momentum states we let $l_1, l_2$
vary within the limits $0-3$. The total number of pseudostates
participating in building the basis states was 20 for each $l$.  To
represent $J = 0,1,2$ singlet states in expansion \eref{exp}, we used
all possible combinations of these pseudostates.  Such a choice gave
us 840 basis states of $S$-symmetry, 1200 basis states of $P$-symmetry
and 1430 states of $D$-symmetry, resulting in a total dimension of the
basis equal to 3470. Issues related to the convergence of the
calculation with respect to the variations of the composition of the
basis set are described in details in Ref.~\cite{tdsep}. A separate
calculation in which we added a subset of 20 pseudostates with $l=4$
produced only a minor change (of an order of a percent) for the
ionization probabilities.

Initial conditions for the solution of TDSE
are determined by solving an eigenvalue
problem using a subset of basis functions of the $S$-symmetry only. 
This
produced the ground state energy of -2.90330~a.u.
We integrate TDSE up to a time $T_1$ when the external field is
switched off. Then we project the solution onto a field-free CCC
wave functions $\Psi({\bm k}_1,{\bm k}_2)$ representing two electrons
in continuum.  Details of the construction of these functions can be
found, for example, in Ref.~\cite{bstel}, or in our earlier paper
\cite{tdsep}.

A set of the final states corresponding to various photo-electron
energies $E_1,E_2$ was prepared. The energies $E_1$ and $E_2$ were
taken on a grid $E_i=1,4,7,10,13,16,19,22,27,40,100,200$
eV. Projection of the solution of the TDSE on the states of this grid
gives us a probability distribution function $p({\bm k}_1,{\bm k}_2)$
of finding the helium atom in a field-free two-electron continuum
state $({\bm k}_1,{\bm k}_2)$ at the time $t=T_1$.

From this probability, we can compute various differential and the total
integrated cross-sections of TPDI. 
The fully resolved, with respect to the photoelectron angles and their
energy, triply differential cross-section (TDCS) is defined as
\begin{equation}
{d\sigma(\omega)\over dE_1 d\Omega_1 d\Omega_2}=
{C\over W q_1q_2 \ \cos^2{\alpha}}
\int
p({\bm k}_1,k_1\tan(\alpha)\ \hat {\bm k}_2) \
k_1 dk_1,
\label{tdcs}
\end{equation}
%
The total integrated cross-section (TICS) is computed as
\begin{equation}
\sigma(\omega)=
{C\over W}\int p({\bm k}_1,{\bm k}_2)
\ d\hat{\bm k}_1 d\hat{\bm k}_2 dk_1 dk_2,
\label{tics}
\end{equation}
Here $\displaystyle W=\int_0^{T_1} F^4_{\rm ac}(t)\ dt$, and
$\displaystyle C=12\pi^2 a_0^4 \tau \omega^2 c^{-2} $ is the
TPDI constant expressed in terms of the speed of light in atomic units
$c\approx 137$, the Bohr radius $a_0= 0.529\times 10^{-8}$~cm and the
atomic unit of time $\tau=2.418\times 10^{-17}$~s.  Momenta $q_1$,
$q_2$ in \Eref{tdcs} are defined on the energy shell: $E_1=q_1^2/2$,
$E-E_1=q_2^2/2$, $\tan{\alpha}=q_2/q_1$, $E$ is the excess energy.

\section{Results.}

There are two TPDI channels with electrons escaping into the $S$ and
$D$ continua.  In the present paper, we are able to report only
results for the $D$-channel as we do not reach satisfactory accuracy
for the $S$-channel.  The reason for this lies in the fact that the
final state CCC wave functions in the $S$-channel are not completely
orthogonal to the ground state wave function. These two sets of
wave functions are obtained using two completely unrelated procedures.
The initial ground $^1S$ state may have, therefore, a nonzero overlap
with the final state CCC wave function which, after propagation in
time, may affect the $S$-channel TPDI results.  Since the $S$-channel
contribution to TPDI is generally a small number, this initial
non-zero overlap can produce considerable inaccuracy in the
calculation of the $S$-wave ionization.

Present results for ionization into the $D$-channel can be utilized in
a two-fold manner. We can either consider them as the exact results
for TPDI in a circular polarized ac field. In this case, only the
$D$-wave contributes as the $S$-wave cannot accommodate two units of
angular momentum projection acquired after absorbing two circularly
polarized photons. Alternatively, we can rely on the fact that the
$S$-wave contribution to TPDI is generally small.  Thus, with some
caution, we can apply the present results to linearly polarized ac
field as well.
To check the accuracy of our method for the $D$-wave, we have in our
disposal the wealth of literature results for the region of photon
energies from 42 to 47~eV, which has been thoroughly studied.

\subsection{Total integrated cross-section}

Before presenting our numerical TICS results across the studied photon
energy range, we wish to outline the procedure we use to attest the
accuracy of our calculation.
Consider the time-evolution of the helium atom in the absence of the
ac external field. This evolution can be presented as a sum
\begin{equation}
\label{evol}
 \Psi(t)=\sum c_k \exp^{-i E_k t} \Psi_k,
\end{equation}
where $\Psi_k$ and $E_k$ are solutions of the eigenvalue problem for
the field-free helium Hamiltonian on the basis \eref{basis}. The
eigenvectors $\Psi_k$ are not strictly orthogonal to the CCC
field-free states.  The overlap of the solution of the TDSE and the
CCC state will therefore contain terms $\displaystyle \sum c_k
\exp^{-i E_k t} \langle \Psi_{\rm CCC}|\Psi_k\rangle$. These terms 
introduce beats in the computed probabilities which may affect the
accuracy of the calculation considerably unless the overlaps $\langle
\Psi_{\rm CCC}|\Psi_k\rangle$ peak in a narrow range of energies
$E_k$.  The magnitude of these beats may serve as an indicator of the
accuracy of the calculation.

\begin{figure}[h]
\epsfxsize=10cm
\epsffile{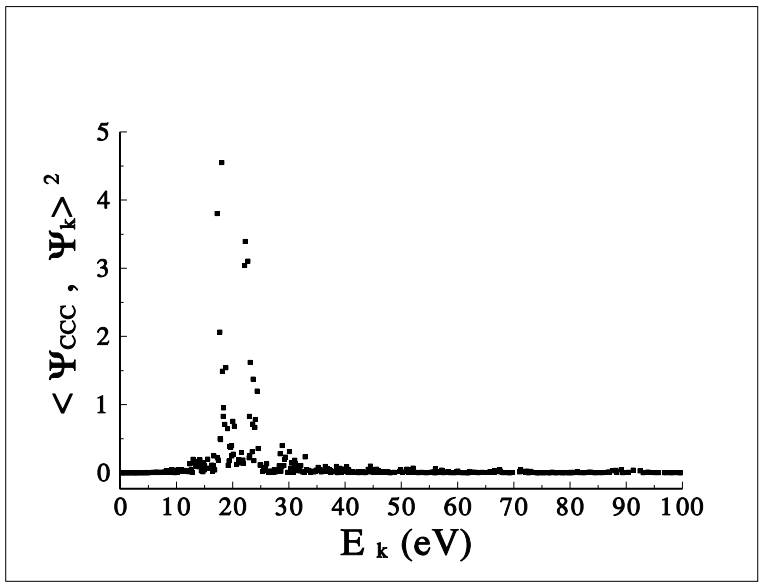}
\caption{
\label{fig1}
Squared overlaps $|\langle \Psi_{\rm CCC}|\Psi_k\rangle|^2$ between
various
$D$-symmetry eiegenfunctions of the
eigenvalue problem for
the field-free helium Hamiltonian on the basis \eref{basis}
and a CCC wave function
at the excess
energy of 20~eV above the double ionization threshold.
}
\end{figure}

This point is illustrated in \Fref{fig1} where we plot the squared
overlaps $|\langle \Psi_{\rm CCC}|\Psi_k\rangle|^2$ between various
$D$-symmetry eiegenfunctions of the 
eigenvalue problem for
the field-free helium Hamiltonian on the basis \eref{basis}
and a final state CCC wave function 
at the excess
energy of 20~eV above the double ionization threshold. We see that
indeed there are only few leading overlaps which peak narrowly around
this energy and other overlaps are insignificant on this scale.

Narrow localization of the overlaps on the energy scale dampens the
beats considerably. This is illustrated in the Table \ref{tab0}
where we present
three sets of TICS computed for several selected photon energies.
These sets are obtained as follows. The first set of TICS (second
column) is computed by overlapping the solution of the TDSE and the
CCC wave functions at the time $T_2=T_1=6T$ when the ac field is switched
off. To obtain the second set of data (third column), we let
the atom evolve freely
for one period after the ac field is switched
off and then the overlaps with the CCC field free states are computed
at the moment $T_2=7T$.  The last set of TICS (the fourth column) is
obtained when the system evolves freely for two periods of the
ac field after it is switched off and the overlaps are computed at the
moment $T_2=8T$.  
As one can see from these data, the beats mentioned above 
lead to variations of TICS of the order of 
20 percent
for the 
photon energy range covered in the Table. We can adopt this figure as 
an estimate of the accuracy of the present calculation.

\begin{table}[h]
\caption{\label{tab0}
TICS (in units of $10^{-52}$ cm$^4$s) obtained for
values of $T_2=6T$, $7T$, and $8T$.}
\begin{ruledtabular}
\begin{tabular}{cccc}
$\omega$  & $6T$  & $7T$ &  $8T$ \\
\hline
\noalign{\smallskip}
42 &  0.500 &  0.443 & 0.506 \\
45 &  0.962 &  0.775 & 0.959 \\
48 &  1.459 &  1.298 & 1.374 \\
50 &  1.646 &  1.768 & 1.629 \\
\noalign{\smallskip}
\end{tabular}
\end{ruledtabular}
\end{table}

For energies outside this range, results are fluctuating much more
and, hence, are considerably less accurate.
This can probably be
explained if we recall the observation we made above about the
nature of the beats
in the computed probabilities. Their magnitude
is determined eventually by the spectrum of
the eigenvalue problem for
the field-free helium Hamiltonian in the basis \eref{basis} and
the set of CCC final state wave functions we use.
Proceeding further into a domain of larger frequencies probably
requires additional tuning of both sets.

\begin{figure}[h]
\epsfxsize=10cm
\epsffile{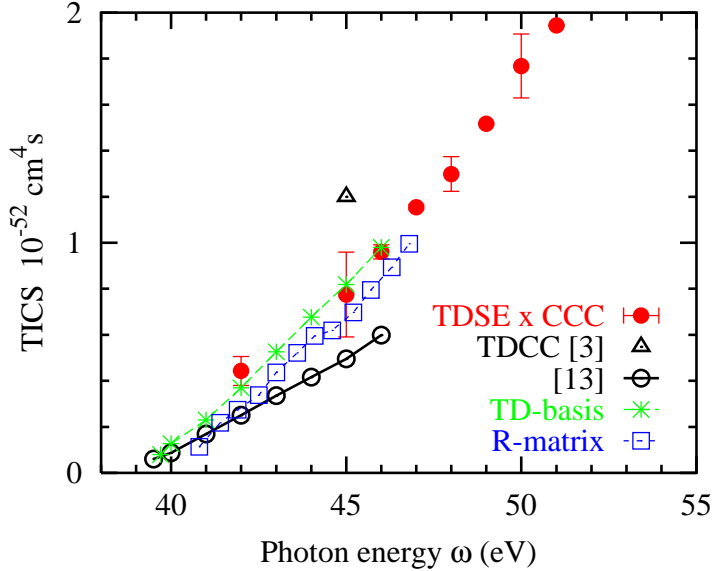}
\caption{
\label{fig2}
Total integrated cross-section of TPDI on He as a function of the
photon energy. Present results obtained by combination of the TDSE and
CCC methods and corresponding to the field intensity of $3.5\times
10^{14}$~W/cm$^2$ are shown by red filled circles. Other calculations
are as follows: TDCC with a $\sin^2$ envelope, $5\times 10^{14}$
W/cm$^2$
\cite{hcc2005} , open circles; TDCC with a ramped pulse, $10^{14}$
W/cm$^2$ \cite{CP02a}, open triangle; TD basis, $10^{14}$ W/cm$^2$
\cite{pbl}, green asterisks; R-matrix, $10^{13}$ W/cm$^2$ \cite{FdH03},
blue open squares.
}
\end{figure}

In \Fref{fig1}, we present our results for TICS in the whole photon
energy range from 42 to 50~eV studied in the paper. The ``error bars''
attached to our data indicate the fluctuation of TICS due to free
propagation beats.  In \Fref{fig1}, we compare the present calculation
with known literature values obtained by the following methods: TDCC
\cite{CP02a,hcc2005},  R-matrix \cite{FdH03} and TD-basis \cite{pbl}.
Within the stated accuracy of 20\%, our results agree with the
R-matrix and TD-basis calculations. The TDCC calculations of
Refs.~\cite{CP02a} and \cite{hcc2005} differ between each other
because two different shapes of the field pulse are utilized in these
works: a constant amplitude pulse which is ramped on and off smoothly
over one field period and a sine squared envelope, respectively.  In
the present calculation we employed a constant amplitude pulse and
therefore our results should be compared with Ref.~\cite{CP02a} which
reported the TICS of $1.2\times 10^{-52}$ cm$^4$s at 45~eV of photon
energy. This is quite close with our result of $9\times 10^{-53}$
cm$^4$s which should further increase when the $S$-wave is accounted
for.

\subsection{Fully differential cross-section}

 

\begin{figure}[h]
\vspace*{5cm}
\hspace*{-6cm}
\epsfxsize=6cm
\epsffile{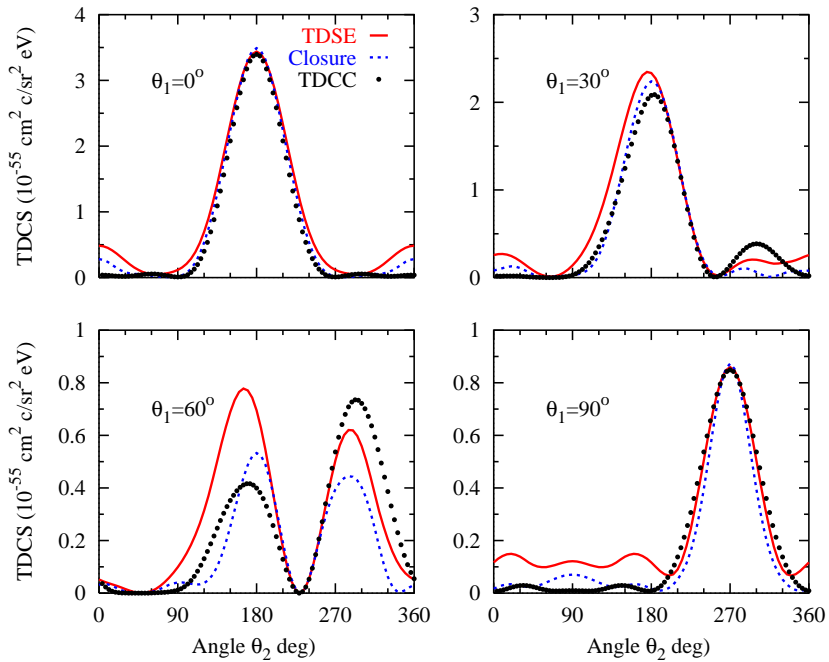}
\caption{
\label{fig3}
 TDCS of He TPDI for the coplanar geometry at $\omega=42$~eV and
 $E_1=E_2=2.5$~eV ($D$-wave contribution only).  The present TDSE
 calculation (divided by 1.3) is shown by the red solid line. The
 earlier CCC calculation in the closure approximation (divided by 1.7)
 is shown by the blue dashed line. The black dots represent the TDCC
 results of Ref.~\cite{hcc2005}.  }
\end{figure}

In \Fref{fig3}, we present our results for the fully resolved TDCS of
TPDI of He at the photon energy of 42~eV and the equal energy sharing
between two photoelectrons $E_1=E_2=2.5$~eV. We adopt the coplanar
geometry in which the momenta of the two photoelectrons and the
polarization vector of light belong to the same plane which is
perpendicular to the propagation direction of the photon. We compare
the present TDSE results with our earlier CCC calculation in the
closure approximation
\cite{KI06}. We also present in the figure 
the TDCC results of \citet{hcc2005} who gave in their work separate
contributions of the $D$ and $S$-waves to TDCS. To make a shape
comparison, we divide the present calculation by the same factor of
1.3 for all fixed electron angles. This factor reflects the difference
in TICS between the two methods. We remind the reader that the TDCC
calculation of   \citet{hcc2005} is performed with a sine squared
envelope and their TICS are smaller than
the present TDSE calculation.
There is a fair shape agreement between the three sets of calculations
except for $\theta_1=60^\circ$ where the relative intensity of two
major peaks is reversed between TDSE and TDCC. The CCC calculation in
the closure approximation is somewhat in between the two other results.

\section{Conclusion.}

In the present work, we studied two-photon double electron ionization
of helium in the range of photon energies from 42 to 50~eV.  
The domain of energies from 42 to 47~eV has been studied extensively
before and there is an abundance of theoretical results in the
literature both for the total and, to lesser extent, differential
cross-sections.  Our present calculations, both for TICS and TDCS,
agree reasonably well with these results. Our TICS values lie on the
higher end of the set of data presented in \Fref{fig2}. As we noted
above, this may be, at least partially, explained by the particular
pulse shape adopted in the present work. More interesting, perhaps, is
the monotonous growth of TICS with the photon energy which we
established for energies below 50 eV.  Most probably, this feature
will be present for any pulse shape.  We may expect some unusual
features to appear in TICS for photon energies approaching the
threshold of sequential TPDI at 54.5~eV.  It was shown in
Refs.~\cite{bach2,bach3} that the spectrum of emitted electrons
undergoes qualitative reconstruction when the new mechanism opens up.
This reconstruction may leave its trace in some additional feature of
TICS. We are going to explore this new regime in the future. We also
intend to resolve the issue of orthogonality and to evaluate the
$S$-wave contribution to TPDI.

The presently analyzed fully differential cross-sections (TDCS) agree
very well between the two CCC calculations: the non-perturbative TDSE
and the perturbative closure. In these two models, we employ the same
CCC final state whereas theoretical description of the field
interaction with the atom is different. The fact that the differential
cross-sections are similar in these two calculations indicates that
the energy and angular correlation in the two-electron continuum is
established as the result of the electron correlation in the final
doubly ionized state. It shows little sensitivity to the precise
mechanism of the atom-field interaction.

\section{Acknowledgements}

We wish to thank James Colgan for supplying the data in numerical
form.  The authors acknowledge support of the Australian Research
Council in the form of the Discovery grant DP0451211. Facilities of
the Australian Partnership for Advanced Computing (APAC) were used.


\end{document}